\documentclass[pra,twocolumn,showpacs,preprintnumbers,amsmath,amssymb]{revtex4-1}
\usepackage{graphicx}
\usepackage{dcolumn}
\usepackage{bm}

\begin{document}

\title{Coherence vortices in one spatial dimension}
\author{Tapio P. Simula and David M. Paganin}
\affiliation{School of Physics, Monash University, Victoria 3800,
Australia} 

\begin{abstract}
Coherence vortices are screw-type topological defects in the phase of Glauber's two-point
degree of quantum coherence, associated with pairs of spatial points at which an ensemble-averaged stochastic
quantum field is uncorrelated. Coherence vortices may be present
in systems whose dimensionality is too low to support spatial
vortices. We exhibit lattices of such quantum-coherence phase 
defects for a one-dimensional model quantum system. We discuss the
physical meaning of coherence vortices and propose how they may
be realized experimentally. \end{abstract} \maketitle

\section{Introduction}

Vortices have fascinated the minds of scientists throughout
history. Beginning with the angular momentum eigenstates of the 
hydrogen atom, quantum-mechanical vortices emerged as key 
entities characterizing quantum liquids such as superconductors
and superfluid helium \cite{Donnelly1991a}. More recently, quantized
vortices have been observed e.g. in Bose-Einstein condensates
\cite{Matthews1999a}, quantum degenerate Fermi gases
\cite{Zwierlein2005a}, and in coherent optical \cite{Tamm1990a}
and acoustic fields \cite{Hefner1999a}. Quantized
vortices, known as coherence vortices, have since been discovered in the cross-spectral density and related
coherence functions of partially-coherent classical-optical fields
\cite{Schouten2003,Gbur2003,GburVisser2006,GburVisserWolf2004,GuGbur2009,Wang2006a,Wang2006b,Marasinghe2010a,Marasinghe2011a}.

Here we show that coherence vortices, which are screw-type singular
phase defects in Glauber's second order degree of quantum
coherence, may exist even in systems with only one spatial
dimension where orbital angular momentum and conventional
quantized vortices cannot be defined. We exhibit a one-dimensional
model system in which decoherence \cite{Zurek2003a,Bennett2000}
is accompanied by a proliferation of quantized phase vortices 
in the coherence function associated with the resulting statistical mixture.
In a space of such low dimensionality, the presence of
a statistical mixture is a necessary condition for the existence
of coherence vortices.  We outline proposals for experimentally
creating and observing coherence vortices and vortex lattices in
spatially one-dimensional systems, and speculate the possible connection
between coherence vortices and decoherence.

\begin{figure*}
\includegraphics[width=1.4\columnwidth]{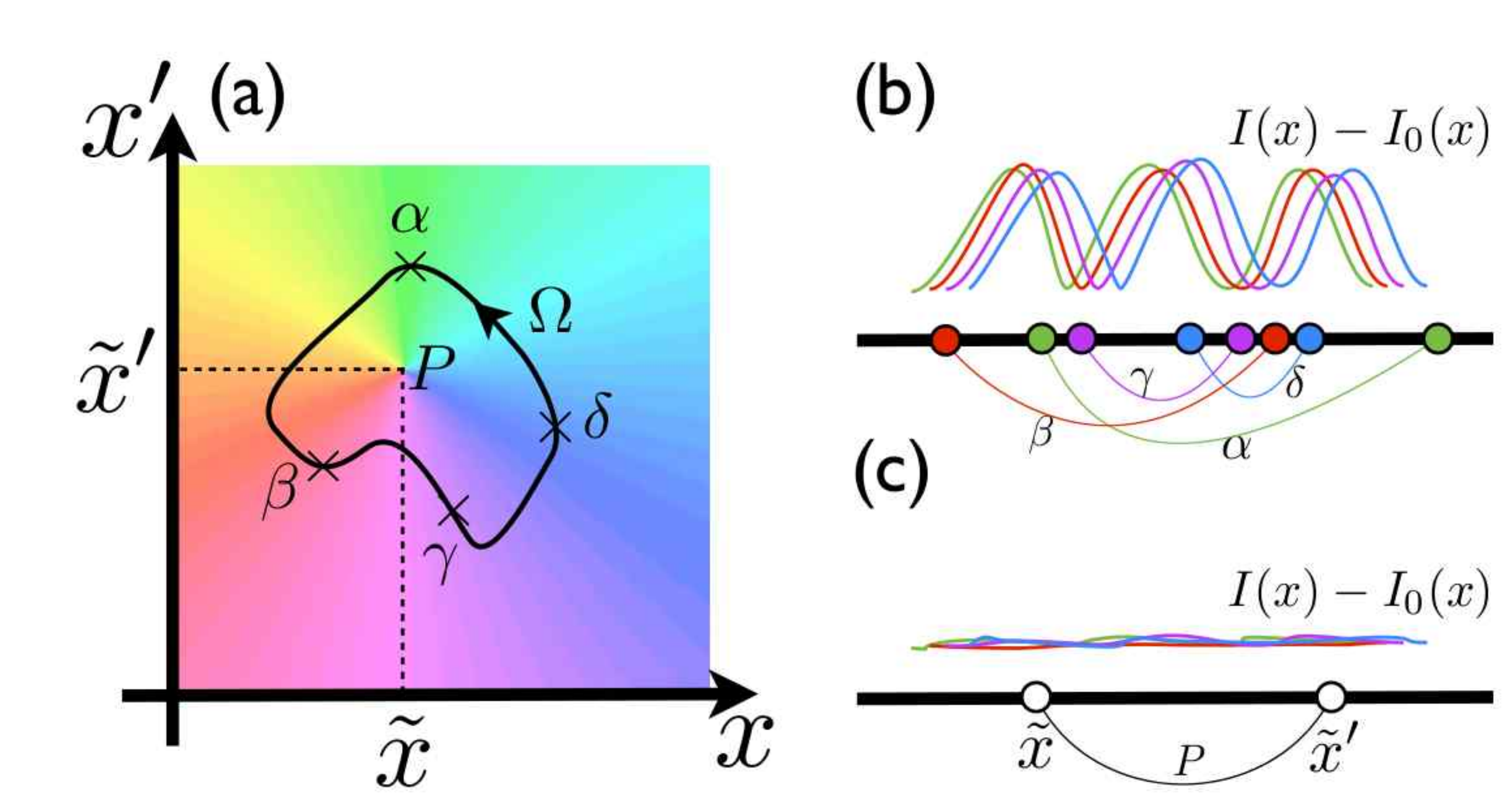}
\caption{(Color online) A coherence vortex as a double-slit interference ratchet.
(a) The presence of a coherence vortex at point $P= (\tilde x, \tilde x')$ is
revealed via non-zero circulation $p$ in
Eq.~(\ref{CoherenceVortexQuantizedCirculation}).
Four coordinate pairs on the contour $\Omega$
are labelled $\alpha, \beta, \gamma$, and $\delta$. 
(b) In general,
non-zero fringe visibility will result should a Young-type interference
experiment be performed by combining the disturbances from the
pair of points corresponding to $\alpha$; the intensity of the
resulting pattern is sketched as a solid (green) line.
Interference patterns are also sketched to correspond to the
coordinate pairs $\beta$ (red), $\gamma$ (magenta) and $\delta$ (blue). As one cycles from
$\alpha\rightarrow\beta\rightarrow\gamma\rightarrow\delta\rightarrow\alpha$,
the corresponding series of Young-type fringes `ratchets' through
one cycle. 
(c) If a Young-type
interference experiment were performed by combining the disturbances from the
pair of points $P = (\tilde x, \tilde x')$, corresponding to the core of the coherence vortex, zero fringe visibility
would result since $g^{(1)}(\tilde x, \tilde x')=0$ (see Eq.~(\ref{GlauberDegreeOfCoherence})).} 
\label{fig1}
\end{figure*}

\section{Quantum vortices}
Vortical flows, associated with velocity fields ${\bf v}({\bf
r},t)$, have non-zero circulation
\begin{equation}\label{SpatialVortexClassicalCirculation}
    \Gamma = \oint_{\Omega}{\bf v}({\bf r},t) \cdot {\bf dl}.
\end{equation}
\noindent Here, $\Omega$ denotes a smooth closed contour
which is traversed once in a positive sense at a fixed time
$t$, ${\bf dl}$ is an infinitesimal arc length vector
 pointing along the direction in which $\Omega$ is
traversed, and $\bf r$ is position. For single-valued differentiable
pure-state complex scalar wave-functions $\psi({\bf r},t)$
with phase $\varphi({\bf r},t)=\arg[\psi({\bf r},t)]$ 
we can define a velocity field
\begin{equation}\label{VelocityQM}
  {\bf v}({\bf r},t) = \frac{\hbar}{m} \nabla \varphi({\bf
r},t),
\end{equation}
where $m$ is the mass of a particle.
The corresponding circulation integral is
quantized in units of $h/m$:
\begin{equation}\label{SpatialVortexQuantizedCirculation2}
    \Gamma = \frac{\hbar}{m} \oint_{\Omega}\nabla\varphi({\bf r},t) \cdot {\bf{dl}}
    = n\frac{h}{m},
\end{equation}
where the integer $n$ is termed the topological charge. This
quantization of circulation is a consequence of the single-valued
and continuous nature of the function $\psi({\bf r},t)$ which may
describe e.g. an optical or a matter-wave field. The integer $n$
denotes the number of circulation quanta and is
related to the orbital angular momentum of the particles 
which is only meaningful in two or more spatial dimensions.

Let us now shift attention, from pure states to mixed states. If
one has a mixed-state wave-function, describing an ergodic
stochastic quantum system via an ensemble of realizations each of
which carry a specified statistical weight, vortical flows may be
studied in terms of the circulation integral in
Eq.(\ref{SpatialVortexClassicalCirculation}) applied to the velocity field
${\bf v}_{\textrm av}({\bf r})=\langle{\bf v}({\bf r},t)\rangle$,
where angular brackets denote ensemble averaging (obtained, for
example, using the density-matrix formalism). We assume there to
be no explicit time-dependence in the external parameters and thus
the time-independence of ${\bf v}_{\textrm av}({\bf r})$ follows
from the ergodicity of the stochastic process describing the
field. The circulation of the averaged velocity field ${\bf
v}_{\textrm av}({\bf r})$ is not quantized in general, even though
circulation of the velocity field is quantized for individual
members of the ensemble.

However, as shown below, quantized circulations exist for {\em
coherence functions} associated with stochastic quantum fields.
Moreover, such quantized coherence circulation may be
meaningfully defined for fields in one spatial dimension, even
though this dimensionality is too low to permit spatial vortical
flows. Lastly, we show that, for the one-dimensional model system 
studied here, a proliferation in the number of coherence vortices and their 
associated coherence-function zeros is associated with the process of decoherence.

\section{Coherence vortices}
For an ergodic stochastic one-dimensional complex quantum field
$\Psi(x,t)$ the normalized two-point equal-time correlation
function is given by the following special case of Glauber's
degree of quantum coherence \cite{Glauber1963a}
\begin{equation}
g^{(1)}(x,x')=\frac{\langle\Psi^\dagger(x,t)\Psi(x',t)\rangle}{\sqrt{|\langle\Psi(x,t)\rangle|^2|\langle\Psi(x',t)\rangle|^2}},
\label{GlauberDegreeOfCoherence}
\end{equation}
where the angular brackets now denote both quantum mechanical
and statistical average. Note that $g^{(1)}(x,x')$ is independent of time, on
account of the assumed ergodicity of the stochastic process.

As a consequence of the single-valued and continuous nature of
each complex field in the ensemble, it follows that
$g^{(1)}(x,x')$ is also a continuous single-valued complex
function of its arguments. Hence, for any simple smooth
positively-traversed closed loop $\Omega$ in the two-dimensional
space coordinatized by $(x,x')$, which is such that
$|g^{(1)}(x,x')|$ is strictly positive for all $(x,x')\in\Omega$,
the coherence-function circulation integral will be given by
\begin{equation}\label{CoherenceVortexQuantizedCirculation}
    \oint_{\Omega}\nabla_{\perp} [\arg g^{(1)}(x,x')] \cdot {\bf dl}_\perp= 2 \pi p,
\end{equation}
where $\perp$ denotes two-dimensional operators and vectors, and
$p$ is an integer denoting topological charge.

Suppose that, for a given closed contour $\Omega$ in the $x$--$x'$
plane shown in Fig.~\ref{fig1}a, $p$ in
Eq.~(\ref{CoherenceVortexQuantizedCirculation}) has a non-zero
value.  Then, there exists at least one coordinate pair $(\tilde
x,\tilde x')$ in the interior of $\Omega$, at which
$g^{(1)}(x,x')$ vanishes. This statement may be proved by
contradiction, as follows: The
continuity of $g^{(1)}(x,x')$ implies that $\arg [g^{(1)}(x,x')]$
will be defined and continuous for every coordinate pair $(x,x')$
inside and on $\Omega$.  Now continuously contract the contour
$\Omega$ to a point, which coincides with some coordinate pair
$(\breve x, \breve x')$ in the the interior of $\Omega$, via the
infinite sequence:
\begin{equation}\label{ContractOmega}
    \Omega \rightarrow \Omega_1 \rightarrow \Omega_2 \rightarrow
    \Omega_3  \cdots \rightarrow (\breve x, \breve
    x')
\end{equation}
which is such that all ``intermediate contours''
$\Omega_1, \Omega_2$ etc. are fully contained within the region
bounded by $\Omega$.  Since $\arg g^{(1)}(x,x')$ can be continuously
contracted to a point without changing
the value of circulation between adjacent members of the
contour sequence in Eq.~(\ref{ContractOmega}), the
non-zero circulation about $\Omega$ (i.e. $2\pi p$, for some
non-zero integer $p$) is equal to the zero circulation about an
infinitesimally-small contour wrapped around $(\breve x, \breve
x')$ (this latter circulation vanishes, since $\arg g^{(1)}(x,x')$
is continuous at and in the infinitesimal vicinity of $(\breve x,
\breve x')$). This logical contradiction implies the falsity of
the initial supposition that $|g^{(1)}(x,x')|>0$ for every
coordinate pair $(x,x')$ inside $\Omega$.  Hence $|g^{(1)}(x,x')|$
vanishes for at least one coordinate pair $(\tilde x,\tilde x')$
inside $\Omega$.

\begin{figure*}
\includegraphics[width=1.2\columnwidth]{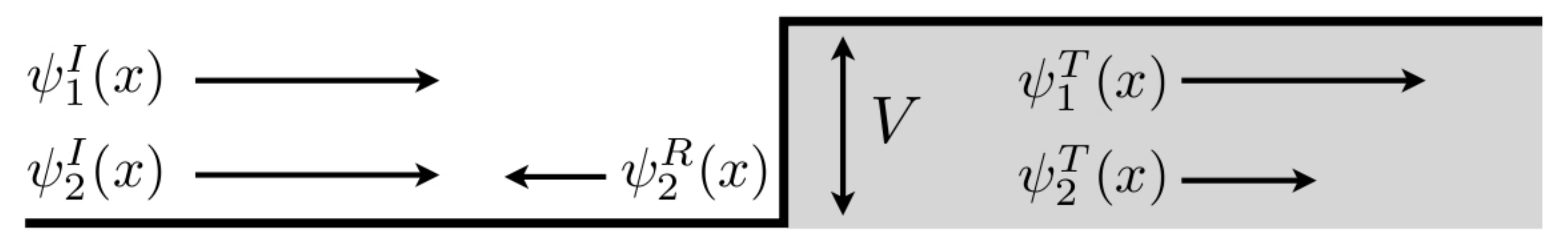}
\caption{Schematic for coherence vortex lattice experiment.
Waves described by $\psi_1(x)$ and $\psi_2(x)$ scatter from a
potential step of height $V$. The potential barrier is transparent
to the first species whereas the second species is both reflected
and transmitted by the potential. } \label{fig4}
\end{figure*}

Non-zero coherence current circulation $2\pi p$ is associated
with quantized coherence vortices, as a low-dimensional special
case of the concept of coherence vortices formulated in the
theory of partially-coherent classical-optics fields, as quantified
through the mutual coherence function or cross-spectral density
\cite{Schouten2003,Gbur2003,GburVisser2006,GburVisserWolf2004,GuGbur2009,Wang2006a,Wang2006b,Marasinghe2010a,Marasinghe2011a}.
In light of the theorem proved above, non-zero $p$ is associated
with at least one pair of points $(\tilde x,\tilde x')$ for which
$g^{(1)}(x,x')$ vanishes.  Notwithstanding that the term ``coherence
vortex'' is employed in the literature, we prefer the term ``decoherence
vortex'' since $(\tilde x,\tilde x')$ is a pair of points at which
the two-point correlation vanishes---an absence of coherence---indicating a
topologically-inevitable zero in the correlation between two
spatially-distinct points in a stochastic quantum field.
Having noted this preference, we will retain the term 
coherence vortex to generically refer to screw-type defects in the phase 
of complex coherence functions such as Glauber's two-point 
correlation function, the mutual coherence function or the cross-spectral density.

The physical meaning of a coherence vortex may be further
clarified with reference to Fig.~\ref{fig1}. The phase of the
complex function $g^{(1)}(x,x')$ is denoted by color in
Fig.~\ref{fig1}(a) where the core of the coherence vortex at
$(\tilde{x},\tilde{x}')$ is labeled by $P$. Four locations
$\alpha, \beta, \gamma$, and $\delta$ corresponding to pairs of
points are marked by crosses along the path $\Omega$, which
encircles the phase singularity at point $P$. Figure \ref{fig1}(b)
shows a schematic of the interference fringes observed in a
Young-type double-slit experiment where the two slits are placed
at points determined by $\alpha, \beta, \gamma$, and $\delta$ and
are illuminated by a spherical wave of light sourced from the
corresponding locations $x$ and $x'$. The background intensity $I_0(x)$
produced by a completely incoherent source field has been
subtracted from the total intensity $I(x)$. The resulting
interference fringes are seen to `ratchet' as the coherence
vortex core is encircled. That is, the troughs and peaks of the
interference pattern move continuously in one direction (within the
overall fringe envelope) when a loop around the coherence vortex
is traversed. If the two slits are placed at the location of the
coherence vortex core $P=(\tilde{x},\tilde{x}')$ the intensity
of the fringes disappears, as illustrated in Fig.~\ref{fig1}(c).

\section{Coherence vortex lattices in one spatial dimension}
\subsection{Experimental systems}
Having discussed the elements of coherence vortices in one
spatial dimension, we now propose experiments to observe them.
Suppose we have a monoenergetic beam of unpolarized photons
propagating from negative infinity $x=-\infty$ through vacuum and
into a block of birefringent material placed at $x=0$ which can be
viewed as a potential step whose height $V$ depends on the state
of the incoming photons. The beam is composed of two unentangled
polarization states of light $|\sigma_+\rangle$ and
$|\sigma_-\rangle$
 and can be described as a statistical mixture of two pure state
 wavefunctions
$\psi_{\sigma_+}(x)$ and $\psi_{\sigma_-}(x)$. Due to the
polarization-dependent propagation in the medium the two components
experience different phase evolution and travel through the material
with different wave vectors. The interference of the different
wave components facilitates the development of coherence vortex
structures. Notice that the light source itself must
{\em not} be strictly coherent, as a necessary condition for the
existence of coherence vortices in one spatial dimension. By performing a set of
interference measurements by varying the locations $x$ and $x'$ of
the interferometer arms the phase map of the two-point coherence
function $g^{(1)}(x,x')$ may be generated, from which the loci of
the coherence vortices may be read off. The photon field could
also be replaced with sound waves to observe acoustic coherence
vortices in one spatial dimension.

Similarly, we may also consider a matter-wave analog of
the electromagnetic field scattering experiment, where the beam of
photons would be replaced by, e.g., an electron beam or a stream of
cold atoms such as those in an atom laser beam sourced from 
a Bose-Einstein condensate. In the case of electrons, the spin-up
$|\uparrow\rangle$ and spin-down $|\downarrow\rangle$ states of
the electrons provide the source of the required mixed states and
an external magnetic field could be used for creating the
polarization dependent potential. In the case of an atomic beam
two or more internal hyperfine spin states would
facilitate the mixing of states and the required potential could
be obtained using optical or magnetic coupling to the atomic
dipole moments. In both cases the measurement of coherence
vortices could be achieved via a two-point interference
experiment. 


\begin{figure*}
\includegraphics[width=1.6\columnwidth]{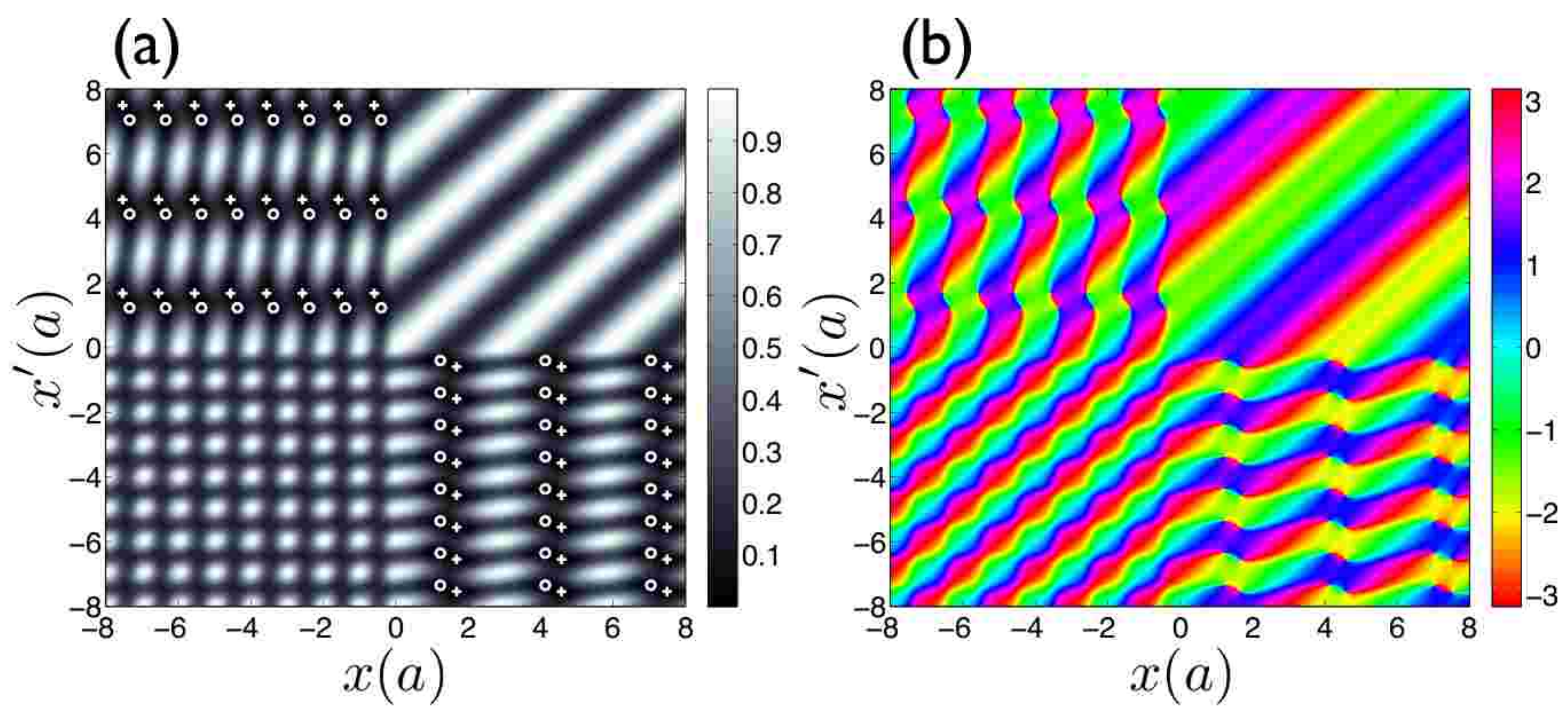}
\caption{(Color online) Stationary interlaced coherence vortex-antivortex dipole lattice.
(a) Coherence density $|g^{(1)}(x,x')|^2$ with the peak density normalized to unity; (b) coherence phase $\arg[g^{(1)}(x,x')]$.
Locations of coherence vortices (coherence antivortices) are marked in
frame (a) by crosses (circles). The unit of length $a = 1/k$ and $V/E=0.99$.}
\label{fig5}
\end{figure*}

\subsection{Theoretical model}
Turning to the theoretical description of the above and related
physical systems, consider a mixed quantum state composed of two 
ensemble members such that
\begin{equation}
g^{(1)}(x,x')=\psi_1^*(x)\psi_1(x') + \psi_2^*(x)\psi_2(x'), 
\label{eqg1}
\end{equation}
less normalization. For incoming plane-wave states, see
Fig.~\ref{fig4}, which scatter from a potential step of height $V$ transparent to 
the first species $\psi_1(x)$, solving the 
Schr\"odinger equation we obtain
\begin{eqnarray}
&&g^{(1)}(x<0,x'>0)= e^{i(-kx+kx')}+T(k,q)e^{i(-kx+qx')}\notag \\
&&+T(k,q)R(k,q)e^{i(kx+qx')},
\end{eqnarray}
where $k$ and $q$ are the wavevectors on each side of the potential step
located at $x=0$ and 
\begin{eqnarray}
T(k,q) &=& 2k/(k+q),\notag \\
R(k,q) &=& (k-q)/(k+q)
\end{eqnarray}
are the transmission
and reflection coefficients, respectively. The energy $E>V$
of the incoming particles of mass $m$ and the height of the potential
step $V$ are linked to the wave vectors by the formulae
\begin{eqnarray}
k^2&=&2mE/\hbar^2,\notag \\
q^2&=&2m(E-V)/\hbar^2. 
\end{eqnarray}

Phase singularities can be produced by an interference of three or more plane waves
\cite{Nicholls1987a,Masaja2001a,Holleran2006a,PaganinBook,Ruben2010a,Ruben2010b}.
Guided by this fact and using complex phasors \cite{Masaja2001a} we solve for
the zeroes of $g^{(1)}(x,x')$ in regions $xx'<0$, to find the locations of the
quantized coherence vortices. We obtain two interlaced lattices
determined by lattice vectors $(x_u,x'_v)$:
\begin{eqnarray}
x_u &=& p\frac{\beta - \alpha}{2k} - u\frac{\pi}{k},\notag\\
     x'_v &=& p\frac{-\alpha}{k-q} +   v\frac{2\pi}{k-q},
            \label{hila}
\end{eqnarray}
where $u$ and $v$ are integers, $x_u<0$ and $x'_v>0$
and the Cartesian shifts are determined by the angles
\begin{eqnarray}
\alpha&=&\pi/p - \arccos (1+T^2-T^2R^2 / 2T),\notag\\
\beta&=&\pi/p + \arccos (1-T^2+T^2R^2 / 2TR).
\end{eqnarray}
These two lattices correspond to, respectively, coherence
vortices and coherence antivortices, determined by
 the sign of the coherence circulation $p=\pm1$.
  In Fig.~\ref{fig5}(a) we have plotted
 the coherence density $|g^{(1)}(x,x')|^2$ scaled by its maximum value. The coherence vortices
 and antivortices have been marked by circles and crosses, respectively,
 at locations determined by Eq.~(\ref{hila}). Figure~\ref{fig5}(b) shows the corresponding
 phase plot $\arg[g^{(1)}(x,x')]$ from which it can be verified that the loci of phase
 singularities correspond to the cores of the coherence vortex lattice.
The Hermitian property of $g^{(1)}(x,x')$ is inherited by the coherence vortex lattice. 
Numerically, it is possible to construct the 
function $g^{(1)}(x,x')$ for an arbitrary potential shape and to find the
resulting coherence vortices and antivortices. For such general 
potential landscapes, the coherence vortices and antivortices 
will not lie on a regular lattice. We emphasize that if the density
matrix for one-dimensional systems can be written in a pure-state form, 
coherence vortices are absent. This follows from the fact that we need a minimum of
three phasors to produce phase singularities via plane-wave interference.

\section{Discussion}
Conventionally, quantized vortices have been studied as phase
singularities in the ``medium" supporting them, e.g.
superconducting order parameter, complex electromagnetic field, or
macroscopic wavefunction describing a matterwave of degenerate
quantum liquids. Here we have considered quantized vortices
emerging in the coherence or correlation function derived from the
quantum states describing the system.
Such phase singularities, known as coherence vortices, are
topologically enforced zeroes in the two-point coherence function.
They appear as a field interference and hence may emerge even when
there are no conventional quantized vortices in the field itself.
In particular, coherence vortices can exist in systems with one
spatial dimension only, as shown here, where conventional
vorticity is manifestly absent. We have described the physics of 
coherence vortices and have
proposed experiments to observe them. We have shown using an
analytically soluble model how a light or matter-wave field
scattering from a step potential creates a stationary coherence
vortex and antivortex lattice. 

Decoherence is considered to be a mechanism for the way 
classical realism emerges from a quantum coherent system 
description in terms of mixed-state density matrices. We speculate
it to be possible that decoherence in a quantum system leaves a topological 
imprint in terms of coherence vortices, which could be used 
for detecting the amount of decoherence in a system.
This possibility is suggested by the fact that, in the model 
one-dimensional system studied here, coherence vortices 
are absent from the pure-state systems and only emerge when 
one has at least the two-member statistical mixture given by Eq.~(\ref{eqg1}).
Further work is also required to clarify whether the theory of coherence vortices may be connected with
the proliferation of thermally activated vortex-antivortex pairs \cite{Simula2006a} and quasi-coherence in 
the observed superfluid to normal transition in cold quasi-two-dimensional 
Bose gases \cite{Hadzibabic2006a,Clade2009a,Tung2010a,Hung2011a}.
Finally, the presence of coherence vortices might
signal some new physical material property in analogy with the way
conventional quantized vortices are inherently linked together
with the phenomena of superfluidity and superconductivity.



\begin{thebibliography}{90}
\bibitem{Donnelly1991a}
R. J. Donnelly, \emph{Quantized vortices in helium II}, Cambridge University Press (Cambridge 1991).
\bibitem{Matthews1999a}
M. R. Matthews, B. P. Anderson, P. C. Haljan, D. S. Hall, C. E. Wieman, and E. A. Cornell, Phys. Rev. Lett. {\bf 83}, 2498 (1999).
\bibitem{Zwierlein2005a}
M. W. Zwierlein, J. R. Abo-Shaeer, A. Schirotzek, C. H. Schunck, and W. Ketterle, Nature {\bf 435}, 1047 (2005).
\bibitem{Tamm1990a}
C. Tamm and C.O. Weiss, J. Opt. Soc. Am. B {\bf 7}, 1034 (1990).
\bibitem{Hefner1999a}
B. T. Hefner and P. L. Marston, J. Acoust. Soc. Am. {\bf 106}, 3313 (1999).
\bibitem{Schouten2003}
H. F. Schouten, G. Gbur, T. D. Visser, and E. Wolf, Opt. Lett. {\bf 28}, 968 (2003).
\bibitem{Gbur2003}
G. Gbur and T. D. Visser, Opt. Commun. {\bf 222}, 117 (2003).
\bibitem{GburVisser2006} 
G. Gbur and T. D. Visser, Opt. Commun. {\bf 259}, 428 (2006).
\bibitem{Wang2006a}
W. Wang, Z. Duan, S. G. Hanson, Y. Miyamoto, and M. Takeda, Phys. Rev. Lett. {\bf 96}, 073902 (2006).
\bibitem{Wang2006b}
W. Wang and M. Takeda, Phys. Rev. Lett. {\bf 96}, 223904 (2006).
\bibitem{GburVisserWolf2004}
G. Gbur, T. D. Visser, and E. Wolf, Opt. Commun. {\bf 239}, 15 (2009).
\bibitem{GuGbur2009}
Y. Gu and G. Gbur, Opt. Commun. {\bf 282}, 709 (2009).
\bibitem{Marasinghe2010a}
M. L. Marasinghe, M. Premaratne, and D. M. Paganin, Opt. Expr. {\bf 18}, 6628 (2010).
\bibitem{Marasinghe2011a}
M. L. Marasinghe, D. M. Paganin and M. Premaratne, Opt. Lett. {\bf 36}, 936 (2011).
\bibitem{Zurek2003a}
W. H. Zurek, 
Rev. Mod. Phys. {\bf 75}, 715 (2003).
\bibitem{Bennett2000}
C. H. Bennett and D. P. DiVincenzo, Nature {\bf 404}, 247 (2000).
\bibitem{Glauber1963a}
R. J. Glauber, Phys. Rev. {\bf 130}, 2529 (1963).
\bibitem{Nicholls1987a}
K. W. Nicholls and J. F. Nye, J. Phys. A: Math. Gen. {\bf 20} 4673 (1987).
\bibitem{Masaja2001a}
J. Masajada and B. Dubik, Opt. Commun. {\bf 198}, 21 (2001).
\bibitem{Holleran2006a}
K. O'Holleran, M. J. Padgett, and M.R. Dennis, Opt. Expr. {\bf 14}, 3039 (2011).
\bibitem{PaganinBook}
D. M. Paganin, \emph{Coherent X-Ray Optics},  Oxford University Press, (Oxford 2006).
\bibitem{Ruben2010a}
G. Ruben, D. M. Paganin, and M. J. Morgan, Phys. Rev. A {\bf 78}, 013631 (2008).
\bibitem{Ruben2010b}
G. Ruben, M. J. Morgan, and D. M. Paganin, Phys. Rev. Lett. {\bf 105}, 220402 (2010).
\bibitem{Simula2006a}
T. P. Simula and P. B. Blakie, Phys. Rev. Lett. {\bf 96}, 020404 (2006).
\bibitem{Hadzibabic2006a}
Z. Hadzibabic, P. Kr\"uger, M. Cheneau, B. Battelier, and J. Dalibard,
Nature {\bf 441}, 1118 (2006). 
\bibitem{Clade2009a}
P. Clad\'e, C. Ryu, A. Ramanathan, K. Helmerson, and W. D. Phillips,
Phys. Rev. Lett. {\bf 102}, 170401 (2009).
\bibitem{Tung2010a}
S. Tung, G. Lamporesi, D. Lobser, L. Xia, and E. A. Cornell,  
Phys. Rev. Lett. {\bf 105}, 230408 (2010). 
\bibitem{Hung2011a}
C.-L. Hung, X. Zhang, N. Gemelke, and C. Chin, Nature {\bf 470}, 236 (2011). 


\end{thebibliography}
\end{document}